\def \SAIT #1 #2 {{\em Mem.\ Soc.\ Astron.\ It.\/} {\bf #1}, #2}
\def \MESS #1 #2 {{\em The Messenger\/} {\bf #1}, #2}
\def \ASTRNACH #1 #2 {{\em Astron. Nach.\/} {\bf #1}, #2}
\def \AAP #1 #2 {{\em Astron. Astrophys.\/} {\bf #1}, #2}
\def \AAL #1 #2 {{\em Astron. Astrophys. Lett.\/} {\bf #1}, L#2}
\def \AAR #1 #2 {{\em Astron. Astrophys. Rev.\/} {\bf #1}, #2}
\def \AAS #1 #2 {{\em Astron. Astrophys. Suppl. Ser.\/} {\bf #1}, #2}
\def \AJ #1 #2 {{\em Astron. J.\/} {\bf #1}, #2}
\def \ANNREV #1 #2 {{\em Ann. Rev. Astron. Astrophys.\/} {\bf #1}, #2}
\def \APJ #1 #2 {{\em Astrophys. J.\/} {\bf #1}, #2}
\def \APJL #1 #2 {{\em Astrophys. J. Lett.\/} {\bf #1}, L#2}
\def \APJS #1 #2 {{\em Astrophys. J. Suppl.\/} {\bf #1}, #2}
\def \APSS #1 #2 {{\em Astrophys. Space Sci.\/} {\bf #1}, #2}
\def \ASR #1 #2 {{\em Adv. Space Res.\/} {\bf #1}, #2}
\def \BAIC #1 #2 {{\em Bull. Astron. Inst. Czechosl.\/} {\bf #1}, #2}
\def \JSQRT #1 #2 {{\em J. Quant. Spectrosc. Radiat. Transfer\/} {\bf #1}, #2}
\def \MN #1 #2 {{\em Mon. Not. R. Astr. Soc.\/} {\bf #1}, #2}
\def \MEM #1 #2 {{\em Mem. R. Astr. Soc.\/} {\bf #1}, #2}
\def \PLR #1 #2 {{\em Phys. Lett. Rev.\/} {\bf #1}, #2}
\def \PASJ #1 #2 {{\em Publ. Astron. Soc. Japan\/} {\bf #1}, #2}
\def \PASP #1 #2 {{\em Publ. Astr. Soc. Pacific\/} {\bf #1}, #2}
\def \NAT #1 #2 {{\em Nature\/} {\bf #1}, #2}

\documentstyle{memsait}
\input epsf.sty
\begin{opening}
\title{SISSA SEARCHING FOR BLAZARS: THE BLEIS PROJECT} % ALL CAPITAL LETTERS PLEASE !!!
\author{Ilaria Cagnoni, Annalisa Celotti, Davide Poccecai}
\institute{International School for Advanced Studies (SISSA-ISAS), Trieste,
Italy}
\date{} % DO NOT INSERT ANY DATE HERE !!!
\end{opening}

\begin{document}

%\oddpagefooter{\sf Mem. S.A.It., Vol. ??, ??}{}{\thepage}
%\evenpagefooter{\thepage}{}{\sf Mem. S.A.It., Vol. ??, ??}
\oddpagefooter{}{}{} % LEAVE AS IT IS !
\evenpagefooter{}{}{} % LEAVE AS IT IS !
\
\vspace{-0.4cm}
\bigskip

\begin{abstract}
We present the aim and the status of the  BLEIS project
currently under development at SISSA.
This project consists in selecting a complete Blazar
sample from deep optical data, down to B=24.6, V=24.4 and I=23.7 (80\%
completness).
The optical images are taken from the ESO Imaging Survey (EIS - Wide), 
a public survey covering $\sim 16$ deg$^2$ of the Southern Emisphere.
This new Blazar sample, thanks to the different energy band 
of the selection and to the faintness of EIS images, 
will be useful not only in understanding the physics 
of such objects, but also in evolutive and cosmological studies.
It will also be a new test for the unified models of AGNs
and will be helpful in making predictions for future deep surveys.
\end{abstract}

\section{Introduction}

The BLEIS (BLazars $+$ EIS) project is a search for blazars from the faint
optical images of the ESO Imaging Survey Wide (EIS Wide).
The aim of the project is to select a complete sample 
to understand the physics and evolutive properties of blazar and to
select interesting sources for VLT.
This will be the first  blazar sample selected from optical images and colors
at such faint fluxes (B=24.6, V=24.4 and I=23.7, 80\% completness).

\section{The EIS Wide}

We used for our project the EIS Wide,  a survey
covering 4 regions of the southern sky for a total of $\sim 16$ deg$^2$ in two/three
colors.
It was planned with the goal of generating a unique publicy available database for the 
ESO community to prepare the widest possible range of programs for the first two
years of operation of VLT (e.g. Renzini \& Da Costa 1997).
Table~1 summarizes the positions and the sky coverage for each filter.\\
EIS wide consists of a mosaic of overlapping EMMI-NTT frames ($9 \times 9$ arcmin) 
with each position of the sky being sampled twice for a total integration time of 300 s
%The average limiting magnitudes at the 80\% completness for the images are
%B=24.6, V=24.4 and I=23.7 
(Da Costa et al. 1999).

\section{Sources selection}

We started our analyis from Patch B, being the region covered with all the 3 filters.
We cross-correlated each EIS Patch B filter catalog with the NRAO VLA Sky Survey
(NVSS, Condon et al. 1998) sources.
The NVSS is a radio survey that covers the sky north of J2000 $\delta = -40^{\circ}$
(82\% of the sky) at 1.4 GHz.
The flux limit is $\sim 2.5$ mJy and the rms uncertainties in right ascension and
declination vary from $\leq 1^{\prime \prime}$ for the sources brighter than 15 mJy to
$\sim 7^{\prime \prime}$ at the survey limit.
For a first preliminary cross-correlation we used a conservative radius of 
$10^{\prime \prime}$ and then performed a more accurate cross-correlation using as 
radii  both the  $1\sigma$ and $2\sigma$ errors associated with the position of each
radio source.
We then excluded all the optical sources fainter than the catalogs 80\% completness
limits and cut at two different radio fluxes: 2.5 mJy (the NVSS limit) and 5 mJy.
We considered only the radio sources with a counterpart in all the 3 filters
whose optical positions lie within $0.8^{\prime \prime}$ one from each other.
%The results of these different selections is presented in Table~2.
%%%%%%%%%%%%%%%%%%%%%%%%%%%%%%%%%%%%%%%%%%%%%%%%%%%%%%%%%%%%%%%%%%%%%%%%%
%\vspace{1cm} %TO ALLOW SUFFICIENT SPACE BETWEEN THE TEXT AND THE FIGURES
%\centerline{\bf Tab. 2 - Number of sources for different selction criteria}

%\begin{table}[h]
%\hspace{3.0cm} %if you want to center your table act on this argument
%\begin{tabular}{|c|c|c|}
%\hline
%Cross-corr. 	&Radio Flux	&Number of\\
%radius$^a$		&mJy		&Source\\
%\hline
%$1\sigma$			&5		&5\\
%$1\sigma$			&2.5		&22\\
%$2\sigma$			&5		&15\\
%$2\sigma$			&2.5		&$\sim 60^b$  \\
%\hline
%\end{tabular}
%\vspace{0.2cm}\\
%{\indent \hspace{0.6in} $^a$ is a multiple of the NVSS positional error for each source\\
%\indent \hspace{0.6in} $^b$ epected number of sources\\
%}
%\end{table}
%%%%%%%%%%%%%%%%%%%%%%%%%%%%%%%%%%%%%%%%%%%%%%%%%%%%%%%%%%%%%%%%%%%%%%%%%%
Cutting at 5 mJy, we find  5 radio sources with at least one optical counterpart
within the $1\sigma$ uncertainty on the radio
position and 15 within $2\sigma$; at 2.5 mJy we have 22 sources within
$1\sigma$ and we expect about 60 sources within $2 \sigma$.
Hereafter we will focus on the 15 sources found within $2\sigma$ 
with a cut at 5mJy.

%%%%%%%%%%%%%%%%%%%%%%%%%%%%%%%%%%%%%%%%%%%%%%%%%%%%%%%%%%%%%%%%%%%%%%%%%
\vspace{0.8cm} %TO ALLOW SUFFICIENT SPACE BETWEEN THE TEXT AND THE FIGURES
\centerline{\bf Tab. 1 - EIS Wide sky coverage}

\begin{table}[h]
\hspace{1.5cm} %if you want to center your table act on this argument
\begin{tabular}{|l|cc|c|c|c|}
\hline
\multicolumn{1}{|c|}{Name} &
\multicolumn{2}{c|}{Center Position} &
\multicolumn{3}{c|}{Area Covered (deg$^2$)}\\
\cline{4-6}

	& &	&B &V &I\\
\hline
Patch A	&22:43:03	&-39:58:30	&--	&1.2	&3.2\\
Patch B	&00:48:22	&-29:31:48	&1.5	&1.5	&1.6\\
Patch C &05:38:24	&-23:51:54	&--	&--	&6.0\\
Patch D &09:51:40	&-21:00:00	&--	&--	&6.0\\
\hline
Total	&		&		&1.5	&2.7	&16.8\\
\hline
\end{tabular}
\end{table}
%%%%%%%%%%%%%%%%%%%%%%%%%%%%%%%%%%%%%%%%%%%%%%%%%%%%%%%%%%%%%%%%%%%%%%%%%%
\vspace{-0.3cm}
\section{Optical classification}

We used the classification based on optical colors presented in Zaggia et al. (1999)
and found that $\sim 3$ sources fall in the region occupied by the high redshift
quasars ($z>3.5$), 5 sources  in the low redshift quasar region ($z<3$) and 5 in an 
unclassified region.
The colors of 3 of these
5 sources are compatible with those of BL Lacertae objects found by Moles et al.
(1985).

\section{$\alpha_{RO}$ distribution}

By extrapolating the NVSS flux (assuming a flat slope to 5~GHz) and using the V magnitudes
we calculated the $\alpha_{RO}$ distribution for our sources and compared it 
to various classes of sources.
Figure~1 shows the BLEIS sources compared to, from top to bottom:\\
 - the X-ray selected BL Lacertae objects of the {\it Einstein}  slew survey (Elvis
et al. 1992);\\
 - the radio selected BL Lacs of the 1Jy sample (Stickel, Meisenheimer \& Kuhr, 1994);\\
 - the radio galaxies of the 3CR sample, whose flux was extrapolated
to 5GHz assuming an average slope for the radio spectrum of $\alpha = 0.7$;\\
 - the FSRQ sample selected by Padovani \& Urry (1992) from the `2Jy sample' of Wall
\& Peacock (1985).\\
As shown in Figure~1, the BLEIS sources properties are compatible with those of
 radio selected BL Lacs and FSRQ.

\begin{figure}
\epsfysize=5.8cm % fix the y-dimension and scales x-dim. to y-dim.
%\epsfxsize=4cm % fix the x-dimension and scales y-dim. to x-dim.
% Feel free to do the choice you prefer but do not exceed the x-dimension
% of the text lines
\hspace{3.5cm}\epsfbox{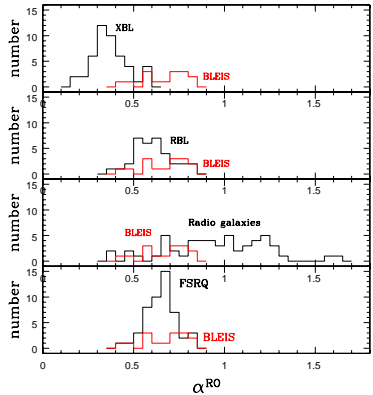} %for centering: act on hspace argument 
\vspace{-0.4cm}
\caption[h]{BLEIS sources $\alpha_{RO}$ distribution compared to those of XBL, RBL, radio
galaxies and FSRQ. See text for details.}
\end{figure}
\vspace{-0.4cm}

\section{Future work}

We plan to extend the analysis to the other patches:
Patch A has V and I data only and there is no NVSS for it, but it will soon be
covered by the Sydney University Molonglo Sky Survey (SUMSS, e.g. Hunstead et al
1998) planned to have the same positional uncertainty and
flux limit of the NVSS.
Patches  C and D have I band images only and NVSS data.
We will ask VLA time to obtain good positions ($\sim 1^{\prime \prime}$),
faint flux limits ($\sim 1$ mJy) and two estimate of the radio flux at different 
wavelengths to be able to select in a confident way fainter sources and have a slope
for their radio spectra.
We will also apply for VLA time for a spectroscopic identification of the interesting
sources and a possible estimate of their redshift.
This would enable us to estimate a luminosity function, a redshift distribution and compare
the results with the predictions of the unified models.
The immediate goal is to determine the counts of blazars at low fluxes and
compare them with the predictions of evolutive and beaming models.

\end{document}